\documentclass[11pt]{article}
\usepackage{latexsym}
\usepackage{epsfig}

\newcommand{\be}{\begin{equation}}
\newcommand{\ee}{\end{equation}}
\newcommand{\ba}{\begin{eqnarray}}
\newcommand{\ea}{\end{eqnarray}}
\newcommand{\baa}{\begin{array}}
\newcommand{\eaa}{\end{array}}

\newcommand{\nr}[1]{(\ref{#1})}

\newcommand{\rmi}[1]{{\mbox{\scriptsize #1}}}
\newcommand{\fr}[2]{{\frac{#1}{#2}\,}}
\newcommand{\mn}{{\mu\nu}}

\def\CL{{\cal L}}

\def\CN{{\cal N}}
\def\gsim{\raise0.3ex\hbox{$>$\kern-0.75em\raise-1.1ex\hbox{$\sim$}}}
\def\lsim{\raise0.3ex\hbox{$<$\kern-0.75em\raise-1.1ex\hbox{$\sim$}}}

\begin{document}

\begin{titlepage}
\begin{flushright}
HIP-2007-27/TH\\
\end{flushright}
\begin{centering}
\vfill

{\Large{\bf Gravity dual of 1+1 dimensional Bjorken expansion}}

\vspace{0.8cm}

\renewcommand{\thefootnote}{\fnsymbol{footnote}}

K. Kajantie$^{\rm a}$\footnote{keijo.kajantie@helsinki.fi},
Jorma Louko$^{\rm b}$\footnote{jorma.louko@nottingham.ac.uk},
T. Tahkokallio$^{\rm a,c}$\footnote{touko.tahkokallio@helsinki.fi}

\setcounter{footnote}{0}

\vspace{0.8cm}

{\em $^{\rm a}$%
Department of Physics, P.O.Box 64, FIN-00014 University of Helsinki,
Finland\\}
{\em $^{\rm b}$%
School of Mathematical Sciences, University of Nottingham,
Nottingham NG7 2RD, UK
\\}
{\em $^{\rm c}$%
Helsinki Institute of Physics, P.O.Box 64, FIN-00014 University of
Helsinki, Finland\\}

\vspace*{0.8cm}

\end{centering}

\noindent
We study the application of AdS/CFT duality to longitudinal boost
invariant Bjorken expansion of QCD matter produced in
ultrarelativistic heavy ion collisions.  As the exact
$(1+4)$-dimensional bulk solutions for the $(1+3)$-dimensional
boundary theory are not known, we investigate in detail the
$(1+1)$-dimensional boundary theory, where the bulk is AdS$_3$
gravity. We find an exact bulk solution, show that this solution
describes part of the spinless Ba\~nados-Teitelboim-Zanelli (BTZ)
black hole with the angular dimension unwrapped, and use the
thermodynamics of the BTZ hole to recover the time-dependent
temperature and entropy density on the boundary. After separating from
the holographic energy-momentum tensor a vacuum contribution, given by
the extremal black hole limit in the bulk, we find that the boundary
fluid is an ideal gas in local thermal equilibrium.
Including angular momentum in the bulk
gives a boundary flow that is boost invariant but has a nonzero
longitudinal velocity with respect to the Bjorken expansion.

\vfill \noindent

%

\vspace*{1cm}

\noindent

Revised September 2007. 

\smallskip 

Published in Phys.\ Rev.\ D {\bf 76}, 106006 (2007). 

\vfill

\end{titlepage}

\section{Introduction}
Collisions of large nuclei at very high energies can be modelled by
taking the transverse size and collision energy to be effectively
infinite, so that the dynamics is invariant under boosts in the
longitudinal direction and translations in the two transverse
directions. When the Minkowski metric is written as
$ds^2=-dt^2+dx^2+dx_2^2+dx_3^2$, where $x$ is the longitudinal
direction, the outcome of the collision takes place in the wedge $t >
|x|$, the natural coordinates are $\tau=\sqrt{t^2-x^2}$ and
$\eta=(1/2)\log[(t+x)/(t-x)]$, the metric reads
$ds^2=-d\tau^2+\tau^2d\eta^2+dx_2^2+dx_3^2$, and the hydrodynamic
variables are independent of $\eta$, $x_2$ and~$x_3$.  Denoting the
shear and bulk viscosities by respectively $\tilde\eta$ and
$\tilde\zeta$, the hydrodynamic equations read $\nabla_\mu T^\mn=0$,
$T^\mn=(\epsilon+p)u^\mu u^\nu+p g^\mn+\Delta T^\mn$, $\Delta
T^\mn=(\fr43\tilde\eta+\tilde\zeta)(g^\mn-u^\mu u^\nu)$,
$\epsilon(T)=3p(T)=3aT^4$, $\tilde\eta=p'(T)/(4\pi)=a T^3/\pi$,
$a=\pi^2N_c^2/8$, and $\tilde\zeta=0$. For the longitudinal similarity
flow,
\be
v(t,x)={x\over t}\equiv\tanh\Theta(\tau,\eta),\qquad
\Theta(\tau,\eta)=\eta,\qquad
u^\mu={x^\mu\over\tau},
\label{simflow}
\ee
these equations can be directly integrated to give
\be
T(t)=\left(T_i+{1\over6\pi\tau_i}\right)
\left({\tau_i\over\tau}\right)^{1/3}-{1\over6\pi\tau},
\label{dissT}
\ee
where $\tau_i$ is normalised by $T(\tau_i)=T_i$ and all the constants
refer to the by now standard predictions for $\CN=4$ supersymmetric
gauge theory \cite{gkp,pss}.  Determining these initial values for
colliding nuclei is an important problem in QCD dynamics \cite{fgm};
very approximately, they are
\cite{ekrt} $T_i,\tau_i=0.5$ GeV, $0.2$ fm/c at $\sqrt s=200$ GeV
(Relativistic Heavy Ion Collider energies) and 
$T_i,\tau_i=1$ GeV, $0.1$ fm/c at $\sqrt s=5500$ GeV 
(CERN LHC energies).

A study of expanding systems in the gauge theory/gravity duality
picture was initiated in
\cite{nastase,ssz,jp}
and continued in several further papers
\cite{jp2,nakamura1,nakamura2, janik1,janik2,kt,kovchegov}. In \cite{jp}
the starting point was to write a candidate
five-dimensional gravity dual of the above collision process
in the form
\be
ds^2 =  {\CL^2 \over z^2} \left[-a(\tau,z) d\tau^2
  + \tau^2 b(\tau,z)d\eta^2+c(\tau,z)(dx_2^2+dx_3^2)+ dz^2   \right]
  \label{jpmetric}
\ee
and then study what constraints five-dimensional Einstein's equations
give for the functions $a$, $b$ and $c$ and to the holographic
energy-momentum tensor computed from them.  In particular, one may
expect to measure the last term in \nr{dissT} and thus independently
check the standard prediction $\tilde\eta/s=1/(4\pi)$. However, since
the holographic energy momentum tensor determines only $\epsilon\sim
T^4$, not $T$ directly, one is not able to measure the last term in
\nr{dissT} independently, and interference terms will appear, even if
one relates energy density directly to the
viscosity~\cite{janik2}. Also, no exact solution for $a$, $b$ and $c$
is known and it is difficult to judge the validity of the several
interesting results
obtained by considering the large-$\tau$ behavior of the
solutions. One may further ask whether the time dependence in
\nr{jpmetric} could be removed by a coordinate
transformation, as the case is for metrics admitting the larger
isometry group $E_3$~\cite{zegers}.  We shall in the following
simplify the problem by neglecting entirely the transverse dimensions
$x_2$ and $x_3$. In the this case Einstein's equations
imply that the bulk geometry is locally AdS$_3$, and the bulk side
becomes exactly solvable.

In case of AdS$_3$ the appropriate 1-brane solution \cite{kraus} in
10d combines $Q_1$ D1-branes in the $x_5$ direction with $Q_5$
D5-branes in the $x^5,...,x^9$ directions.  The $x^6,...,x^9$
directions are compactified on a 4-torus $T_4$ with $V_4\sim
\alpha'^2$ and the $x^5$ is taken to be of length
$L\gg\sqrt{\alpha'}$. When the $x^1,...,x^4$ are written in spherical
coordinates with $ds^2=dr^2+r^2d\Omega_3^2$, the $0,5,r$ coordinates
in the $r\to0$ limit give us AdS$_3$ and the whole structure is
AdS$_3\times$S$_3\times T_4$.  The boundary theory dual to string
theory in this background is a 2d conformal field theory in $x^0,x^5$
with $4Q_1Q_5$ bosons and an equal number of fermions. We shall
neglect the dilaton and the 3-form field strength and determine the
metric by solving AdS$_3$ gravity equations. Several different
coordinate systems are studied. The holographic energy momentum tensor
is determined. We observe that the energy density computed in this way
is exactly that of an ideal gas of $4Q_1Q_5$ bosons
and fermions, the factor 3/4 observed for the usual AdS$_5$
case is 1 now.

The differences in the application of AdS/CFT duality to
AdS$_5\times$S$_5$ and AdS$_3\times$S$_3\times T_4$ are manifest in
the relation between string theory and supergravity background
parameters. For AdS$_5\times$S$_5$ we have
\be
\CL^4= 4\pi g_sN_c\alpha'^2=g_\rmi{YM}^2N_c\alpha'^2,\qquad
{\CL^3\over G_5}={2N_c^2\over\pi},
\ee
where the string coupling $g_s$ is constant since the dilaton is
a constant. Thus the AdS$_5$ radius and the string tension are simply
related via the coupling constant $g_\rmi{YM}$ of the boundary theory,
${\cal N}=4$ super-Yang-Mills (SYM). For AdS$_3\times$S$_3\times T_4$,
\be
\CL^4=g_s^2{16\pi^4\alpha'^2\over V_4}Q_1Q_5\alpha'^2,\qquad
{\CL\over G_3}=4Q_1Q_5.
\label{g3}
\ee
Thus the relation between the string tension and AdS$_3$ depends on
the compactification volume $V_4$, which is not an experimental
number. However, for both cases the dimensionless relation between
$\CL$ and the Newton constant is very simple, $\sim$ number of degrees
of freedom.

While the integration of hydrodynamical equations $\nabla_\mu T^\mn=0$
is trivial for given initial conditions, the real problem is in the
determination of initial conditions. For heavy ion collisions these
will depend on the atomic number $A$ of the colliding nuclei and the
collision energy \cite{fgm,ekrt}.  AdS/CFT even in the well controlled
case of ${\cal N}=4$ SYM permits the determination of vacuum
expectation values of operators with fields integrated over
$-\infty<t<\infty$, but the application to path integrals with fields
starting at some $t_i$ seems to be an open question.

\section{The bulk solution}

The local properties of solutions to AdS$_3$ gravity equations are
well known \cite{henneaux,carlip}, but since we aim at a solution with
a specific structure, the derivation gives some insight.  The AdS$_3$
action with the cosmological constant $\Lambda=1/\CL^2$ and the 3d
Einstein equations are
\be
S={1\over16\pi G_3}\int d^3x\sqrt{-g}\left(R+{2\over\CL^2}\right),
\label{einstein-action}
\ee
\be
R_{ab}-\frac{1}{2}R\, g_{ab}-{1\over\CL^2}g_{ab}=0,
\label{einstein-equations}
\ee
which further imply
\be
R=-{6\over\CL^2},\qquad R_{ab}=-{2\over\CL^2}\,g_{ab},\qquad
R_{abcd}={1\over\CL^2}(g_{ad}g_{bc}-g_{ac}g_{bd}),
\ee
\be
R^{ab}R_{ab}=R^{abcd}R_{abcd}={12\over\CL^4}.
\label{invariants}
\ee
As the $(2+1)$-dimensional analogy of~\nr{jpmetric}, we adopt the
ansatz\footnote{One can also start from an ansatz with
$\eta$-dependence, i.e.~$a(\tau,z)\to a(\tau,\eta,z)$ and
$b(\tau,z)\to b(\tau,\eta,z)$, but if one wants the diag$(-1,\tau^2)$
boundary metric one can show that all $\eta$ dependence dies away. The
absence of cross-terms in the ansatz is crucial for this result.}
\be
ds^2=\frac{\CL^2}{z^2}\left[-a^2(\tau,z)
d\tau^2+b^2(\tau,z) d\eta^2+dz^2\right] ,
\label{ansatz}
\ee
where $z>0$ and $-\infty < \eta < \infty$.
The nontrivial components of Einstein's equations then yield the
following set of equations:
\ba
\tau\tau:\quad&&\partial_z a-z\partial_z^2a=0,
\label{eq:tautau}
\\
\eta\eta:\quad&&\partial_z b-z\partial_z^2b=0,
\label{eq:etaeta}
\\
\tau z:\quad&&\partial_z a\,\partial_\tau b
=a\,\partial_z\partial_\tau b,
\label{eq:tauz}
\\
zz:\quad&&-a^3\partial_zb+a^2\partial_za(-b+z\partial_zb)+z\partial_\tau a
\partial_\tau b-z\,a\,\partial_\tau^2b=0.
\label{eq:zz}
\ea

We look for solutions that have at $z\to0$ the asymptotically AdS form
\cite{skenderis}
\be
ds^2={\CL^2\over z^2} [ g_\mn dx^\mu dx^\nu+dz^2 ],
\label{AdS-form-in-z}
\ee
where the two-dimensional metric $g_\mn$ has the small $z$ expansion
\be
g_\mn(\tau,z)=g^{(0)}_\mn(\tau)
+g^{(2)}_\mn(\tau)z^2+g^{(4)}_\mn(\tau)z^4+ \ldots ,
\label{gexp}
\ee
and the conformal boundary metric $g^{(0)}_\mn$ is the Milne
metric~\cite{ellis-williams}
\be
g^{(0)}_\mn dx^\mu dx^\nu
= - d\tau^2 + \tau^2 d\eta^2 ,
\label{milne-metric}
\ee
with $0 <\tau < \infty$.
Equations \nr{eq:tautau} and
\nr{eq:etaeta} integrate immediately to
\ba
a(\tau,z)&=& a_1(\tau)z^2+a_2(\tau) , \\
b(\tau,z)&=&b_1(\tau)z^2+b_2(\tau).
\ea
Matching to the boundary metric \nr{milne-metric} gives
$a_2^2(\tau)=1$ and $b_2^2(\tau)=\tau^2$. Equations
\nr{eq:tauz}
and
\nr{eq:zz}
then reduce to a form from which $a_1$ and $b_1$ can be found by
elementary integration. The general solution can be written as
\be
ds^2=\frac{\CL^2}{z^2}
\left[-\left(1-\frac{(M-1)z^2}{4\tau^2}\right)^2
d\tau^2+ \left(1+\frac{(M-1)z^2}{4\tau^2}\right)^2
\tau^2 d\eta^2+dz^2\right] ,
\label{t-solution-M}
\ee
where the constant $M$ may take any real value.

We shall from now on assume $M$ to be positive. It will be explained
at the end of section \ref{sec:BTZ-spinless} why this is the only case
in which we can recover interesting thermodynamics.

\section{Relation to the AdS$_3$ black hole}
\label{sec:BTZ-spinless}

The metric \nr{t-solution-M} is explicitly time dependent, and it has
a coordinate singularity at $\tau^2 = |M-1|z^2/4$. As a solution to
Einstein's equations~\nr{einstein-equations}, the metric must however
be locally AdS$_3$~\cite{henneaux,carlip}. We shall now show that the
metric covers part of the spinless Ba\~nados-Teitelboim-Zanelli (BTZ)
black hole whose angular dimension has been unwrapped, and the
coordinate singularity resides in the white hole region of this
spacetime.

\begin{figure}[!tb]
\begin{center}

\vspace{-0.8cm}
\includegraphics[width=0.6\textwidth]{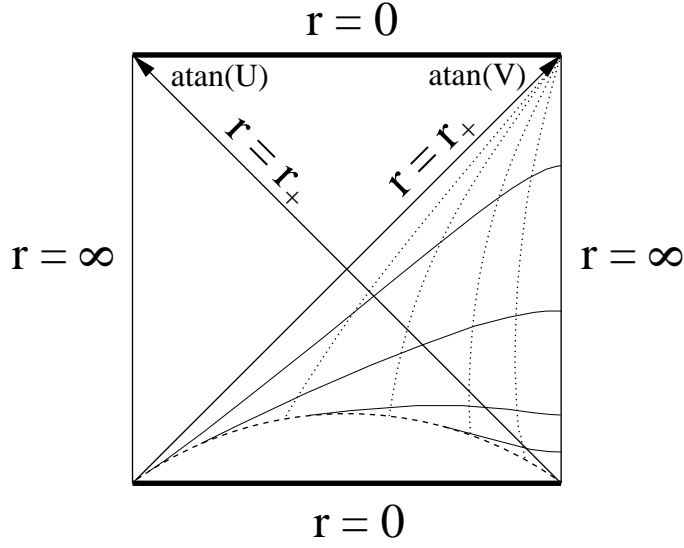}
\end{center}
\vspace{-0.8cm}
\caption{\small
The conformal diagram of the spinless nonextremal BTZ black-and-white
hole \nr{spinlessBTZ-kruskal} \cite{henneaux,carlip}. The coordinate
$\eta$ is suppressed, the coordinates of the diagram are $p = \arctan
U$ and $q = \arctan V$ with $|p+q| < \pi/2$ and $|p-q| <
\pi/2$, and the $p$-axis (respectively $q$-axis) is tilted 45 degrees
to the left (right) from the vertical. The Killing horizon of the
Killing vector $\partial_t = (\sqrt{M}/\CL) ( V\partial_V -
U\partial_U )$ is at $pq=0$, where $r = r_+ = \CL\sqrt{M}$. The
coordinate grid shows the curves of constant $\tau$ (solid curves) and
the curves of constant $z$ (dotted curves) in the region $0 < z <
v\tau$, where $v$ is given by
\nr{v-def} and $M>1$ is assumed.
The dashed line is the coordinate singularity
$z = v\tau$, $r=\CL\sqrt{M-1}$.  On a given curve of constant~$\tau$,
$z$~takes the values $0<z<v\tau$, increasing from right to left, and
on a given curve of constant~$z$, $\tau$~takes the values
$z/v<\tau<\infty$, increasing bottom to top.}
\label{BTZnospinfigure}
\end{figure}

Introducing the coordinates $(U,V)$ by
\ba
U &=&
-
\left(
\frac{2\tau -  \left(\sqrt{M}-1\right)z}
{2\tau +  \left(\sqrt{M}+1\right)z}
\right)
\left(\frac{\tau}{\CL}\right)^{-\sqrt{M}} ,
\nonumber
\\
\noalign{\smallskip}
V &=&
\left(
\frac{2\tau -  \left(\sqrt{M}+1\right)z}
{2\tau +  \left(\sqrt{M}-1\right)z}
\right)
\left(\frac{\tau}{\CL}\right)^{\sqrt{M}} ,
\label{tau-z-to-kruskal}
\ea
the metric \nr{t-solution-M} takes the form
\be
ds^2
=
\CL^2
\left[
- \frac{ 4 \, dU\,dV}{{(1+UV)}^2}
+ M
\left(
\frac{1-UV}{1+UV}
\right)^2
d\eta^2
\right]
\ \ .
\label{spinlessBTZ-kruskal}
\ee
Suppose for the moment that $\eta$ were periodic with period
$2\pi$. The metric
\nr{spinlessBTZ-kruskal} is then the
spinless nonextremal BTZ black hole, and the global Kruskal-type null
coordinates $(U,V)$ have the range $-1 < UV < 1$
\cite{henneaux,carlip}. The left and right infinities are at $UV
\to -1$, the Killing horizon of the Killing vector $V\partial_V -
U\partial_U$ is at $UV=0$, and the black and white hole orbifold
singularities are at $UV \to 1$. The conformal diagram is shown in
Figure~\ref{BTZnospinfigure}.  In each of the quadrants in which
$UV\ne0$, the coordinate transformation
\be
r
=
\CL \sqrt{M} \, \left(\frac{1-UV}{1+UV}\right) ,
\qquad
t
=
\frac{\CL}{2\sqrt{M}} \ln \! \left|\frac{V}{U}\right| ,
\label{kruskal-to-schw}
\ee
brings the metric to the Schwarzschild-like form
\begin{equation}
ds^2 =
- \left( {r^2\over\CL^2} - M \right) dt^2+\frac{dr^2}{r^2/\CL^2 - M}+
r^2 d\eta^2 ,
\label{standard}
\end{equation}
in which $\partial_t = (\sqrt{M}/\CL) ( V\partial_V - U\partial_U )$,
and the Killing horizon of $\partial_t$ is at $r =\CL\sqrt{M}$.
The Arnowitt-Deser-Misner (ADM) 
energy equals $M/(8G_3)$, where the factor $1/(8G_3)$ comes
from our normalisation of the Einstein action~\nr{einstein-action}.

In our spacetime $\eta$ is not periodic, and the singularities of
\nr{spinlessBTZ-kruskal} at $UV\to1$ are just singularities of
the coordinate system $(U,V,\eta)$ on the AdS$_3$ hyperboloid
\cite{henneaux,carlip}. What is important for us is that the
observations about the Killing vector $\partial_t$ remain valid even
when $\eta$ is not periodic. We shall continue to describe the metric
\nr{spinlessBTZ-kruskal} in the
black hole terminology even for nonperiodic~$\eta$, trusting that no
ambiguity will ensue.

The region of the BTZ hole \nr{spinlessBTZ-kruskal} covered by
\nr{t-solution-M} depends on whether $0<M\le1$ or
$1<M<\infty$. Suppose first that $1<M<\infty$. The metric
\nr{t-solution-M} is then
regular for $z < v \tau$ and $v \tau<z$, where
\begin{equation}
v = \frac{2}{\sqrt{M-1}} .
\label{v-def}
\end{equation}
Examination of \nr{tau-z-to-kruskal} shows that the region $z < v
\tau$ covers in
\nr{spinlessBTZ-kruskal} the region
\be
U<0 ,
\qquad
V >
\left(
\frac{\sqrt{M} - \sqrt{M-1}}
{\sqrt{M} + \sqrt{M-1}}
\right)
\left(\frac{1}{U}\right) ,
\label{region-in-kruskal}
\ee
and the coordinate singularity at $z = v\tau$ is on the spacelike
surface of constant $UV$ where the latter inequality in
\nr{region-in-kruskal} becomes saturated. The curves of constant
$\tau$, the curves of constant $z$ and the coordinate singularity are
shown in Figure~\ref{BTZnospinfigure}. If we take $\tau$ to increase
to the future, so that $U$ and $V$ increase to the future, the region
$z < v\tau$ thus comprises one exterior region, where
$\CL\sqrt{M}<r<\infty$, and the part $\CL\sqrt{M-1}<r<\CL\sqrt{M}$ of
the white hole interior region. The coordinate singularity is on the
spacelike surface $r=\CL\sqrt{M-1}$ in the white hole interior.  The
parameter $v$ has an interpretation as the coordinate velocity of the
coordinate singularity, but this coordinate singularity is neither an
event horizon, apparent horizon nor a dynamical horizon
\cite{hawking-ellis,ashtekar-dynamical}, and would not be even if
$\eta$ were made periodic to satisfy certain technical conditions in
the definitions of these horizons. The coordinate singularity arises
just because the surfaces of constant $\tau$ become parallel to the
surfaces of constant~$z/\tau$.

Similar considerations hold for the region $v \tau < z$, and the
region covered in \nr{spinlessBTZ-kruskal} is obtained by
interchanging $U$ and $V$ in~\nr{region-in-kruskal}. A~qualitative
picture of the curves of constant $\tau$ and the curves of constant
$z$ is obtained by a left-right interchange in
Figure~\ref{BTZnospinfigure}.  As we are interested in the conformal
boundary where $z\to0$ with fixed $\tau$ and~$\eta$, the region of
interest for us is however $z < v \tau$.


Suppose then that $0<M \le1$. The coordinate singularity is now at
$\tau = \frac12 \sqrt{1-M} \, z$,
and in
the conformal diagram of Figure \ref{BTZnospinfigure} it is at the
white hole singularity. The region
$\frac12 \sqrt{1-M} \, z < \tau$, which
reaches the conformal boundary at $z\to0$, thus covers in
\nr{spinlessBTZ-kruskal} one exterior and all of the white hole
interior.

For use in section~\ref{sec:thermodynamics}, we record here some
thermodynamical properties of the BTZ metric. For quantum fields in
AdS$_3$, an AdS-invariant vacuum state induces in the metric
\nr{standard} a state that is thermal with respect to the Killing
vector~$\partial_t$, in the BTZ temperature
\cite{henneaux,carlip}
\be
T_{\mathrm{BTZ}}
=
\frac{\sqrt{M}}{2\pi\CL} .
\label{T-BTZ}
\ee
The ADM energy $\Delta E$ and the Bekenstein-Hawking entropy
$\Delta S_{\mathrm{BTZ}}$ for an interval $\Delta \eta$ are
\be
{\Delta E\over \Delta \eta}={M\over16\pi G_3},\qquad
{\Delta S_{\mathrm{BTZ}}\over\CL \Delta \eta}=
\frac{\sqrt{M}}{4 G_3} .
\label{s-BTZ}
\ee
These formulas originally arose in the context in which $\eta$ has
period $2\pi$ and the bulk spacetime is the BTZ black hole,
but they remain valid also for nonperiodic $\eta$ if one maintains
that all thermodynamical quantities should be defined so that they are
invariant under translations in~$\eta$. It is this assumption of
translational invariance that forces one to regard $\partial_t$ in
\nr{standard} as the physically relevant time translation Killing
vector, as the only Killing vectors that commute with translations in
$\eta$ are $\partial_\eta$ and~$\partial_t$.
For the BTZ hole the translational invariance comes from
the periodicity of~$\eta$, while for us it is motivated by the aim to
model a boost-invariant flow on the conformal boundary.

For use in section \ref{sec:thermodynamics}, we also recall that the
coordinate transformation
\be
r = \frac{\CL^2}{z} \left( 1 + \frac{M z^2}{4\CL^2} \right)
\ee
(where $z$ is not the same as that in~\nr{t-solution-M}) brings the
metric \nr{standard} to the asymptotically AdS standard form of
\nr{AdS-form-in-z} and~\nr{gexp},
\be
ds^2 =
\frac{\CL^2}{z^2}
\left[
- \left( 1 - \frac{Mz^2}{4\CL^2} \right)^2 dt^2
+ \left( 1 + \frac{Mz^2}{4\CL^2} \right)^2 \CL^2 d\eta^2+dz^2
\right] .
\label{eq:static-metric}
\ee

To end this section, recall that we have throughout assumed
$M>0$. When $M\le0$, the transformation \nr{tau-z-to-kruskal} becomes
ill defined, but it can be verified that the transformation from
\nr{t-solution-M} directly to
\nr{standard} remains well defined. For $M=0$, \nr{standard} is the
extremal spinless BTZ hole with the angular direction unwrapped, and
the Killing horizon of $\partial_t$ is degenerate and has vanishing
temperature. For $M<0$, \nr{standard} has a massive point particle at
$r=0$ with the angular direction unwrapped, and $\partial_t$ does not
have a Killing horizon. It is therefore only with $M>0$, or in the
limiting case $M=0$, that we can associate to the metric \nr{standard}
a temperature and an entropy.

\section{Energy-momentum and thermodynamics}
\label{sec:thermodynamics}

Our metric \nr{t-solution-M} has the asymptotically AdS form of
\nr{AdS-form-in-z} and~\nr{gexp},
where the conformal boundary metric is the Milne
metric~\nr{milne-metric}. On the conformal boundary of AdS$_3$, this
Milne universe covers the diamond in which the region $U<0$, $V>0$ of
\nr{spinlessBTZ-kruskal} meets the conformal
boundary~\cite{aminneborg}.

The boundary energy-momentum tensor can be calculated using
holographic renormalization equations. In our case of a two-dimensional
boundary, the formula is \cite{skenderis}
\be
T_{\mu\nu}=\frac{\CL}{8\pi G_3}
[g_{\mu\nu}^{(2)}-g_{\mu\nu}^{(0)}\textrm{Tr}(g_{\mu\nu}^{(2)})].
\label{skende}
\ee
Working in the co-moving Milne coordinates $(\tau,\eta)$
of~\nr{milne-metric}, we find
\be
T_{\mu \nu} = {\CL (M-1) \over 16\pi G_3 }
\left(
\begin{array}{cc}
\tau^{-2} & 0 \\
\noalign{\smallskip}
0 & 1
\end{array}
\right) .
\label{raw-T}
\ee
While \nr{raw-T} should now contain the energy-momentum tensor of the
fluid whose thermodynamics we wish to recover, it could also contain a
vacuum energy contribution. As the temperature \nr{T-BTZ} of the bulk
black hole vanishes in the limit $M\to0$, we expect this to be the
limit in which the boundary fluid disappears. We therefore split the
total energy-momentum tensor as $T_{\mu \nu} = T_{\mu
\nu}^{(\mathrm{fluid})} + T_{\mu
\nu}^{(\mathrm{vac})}$, where
\ba
T_{\mu \nu}^{(\mathrm{fluid})}
&=& {\CL M \over 16\pi G_3 }
\left(
\begin{array}{cc}
\tau^{-2} & 0 \\
\noalign{\smallskip}
0 & 1
\end{array}
\right) ,
\label{fluid-T}
\\
\noalign{\smallskip}
T_{\mu \nu}^{(\mathrm{vac})}
&=& - {\CL \over 16\pi G_3 }
\left(
\begin{array}{cc}
\tau^{-2} & 0 \\
\noalign{\smallskip}
0 & 1
\end{array}
\right) ,
\label{vac-T}
\ea
and we interpret $T_{\mu \nu}^{(\mathrm{fluid})}$ and $T_{\mu
\nu}^{(\mathrm{vac})}$ as respectively the fluid and vacuum
contributions.

As $T_{\mu \nu}^{(\mathrm{vac})}$ is not proportional to the boundary
metric, it cannot be the energy-momentum tensor in any
Poincare-invariant vacuum state, not even after any Poincare-invariant
renormalisation. It is however invariant under the boosts generated
by~$\partial_\eta$, and it is proportional to the energy-momentum
tensor of a massless scalar field in the conformal vacuum of the Milne
universe~\cite{byd}. The boundary vacuum state from which $T_{\mu
\nu}^{(\mathrm{vac})}$ arises is therefore unconventional from the ion
collision perspective but instead adapted to the conformal invariance
of the gauge/gravity duality.

The fluid contribution to the energy-momentum tensor takes the perfect
fluid form,
\be
T_{\mu\nu}^{(\mathrm{fluid})}
= (\epsilon + p) u_\mu u_\nu + p g^{(0)}_\mn ,
\label{milne-T-perffluid}
\ee
where the fluid's normalised velocity vector is $u^\mu = (1,0)$ and
the energy density $\epsilon$ and pressure $p$ are given by
\be
\epsilon(\tau)=p(\tau)
= {\CL\over 16\pi G_3}\,{M\over \tau^2} .
\label{eps}
\ee
The fluid is thus comoving in the Milne universe, following the
$(1+1)$-dimensional version of the
Bjorken flow~\nr{simflow}.

This was straightforward; the real problem is to define for this time
dependent situation a time dependent temperature $T(\tau)$ and entropy
density $s(\tau)$ so that $s(T(\tau))=dp(T)/dT$ with $p$ defined
in~\nr{eps}. We are not aware of a method that could be justified with
the same rigor as in the static case, but we now present two
independent arguments, both of which lead to the same result.

\begin{itemize}
\item
The first argument starts from the entropy density. Recall that in the
static case the entropy density was given by \nr{s-BTZ} with $\CL
d\eta$ as the longitudinal volume element. For the time dependent
expanding case the longitudinal volume element is, instead, $\tau
d\eta$. Thus the proper entropy density and temperature (the latter
from $s\sim T$ for an ideal gas in $1+1$ dimensions) are
\be
s(\tau)
={\Delta S\over \tau \Delta\eta}
=\frac{\sqrt{M}}{4 G_3}{\CL\over\tau},
\qquad
T(\tau)={\sqrt{M}\over2\pi\tau}.
\label{T-s}
\ee
Inserting
\nr{T-s} in~\nr{eps}, we
find
\be
\epsilon(\tau)=p(\tau)
={\pi\CL\over 4 G_3} T^2(\tau),
\quad
s(\tau)={\pi\CL\over 2 G_3}T(\tau).
\ee
Note that this result satisfies the proper thermodynamic
relation $s(T)=p'(T)$.

\item
The second argument starts from the temperature. The conformal
boundary metric in \nr{eq:static-metric} is the Minkowski metric,
$ds^2_{\mathrm{CBTZ}} = - dt^2 + \CL^2 d\eta^2$, and it follows from
\nr{tau-z-to-kruskal}
and
\nr{kruskal-to-schw} that
$\tau/\CL = e^{t/\CL}$.
The boundary metrics in \nr{t-solution-M} and
\nr{eq:static-metric}
are thus related by the time-dependent
conformal scaling
\be
ds^2_{\mathrm{Milne}}
= -d\tau^2 + \tau^2 d\eta^2
=
{(\tau/\CL)}^2
\left( - dt^2 + \CL^2 d\eta^2 \right)
= {(\tau/\CL)}^2 \,
ds^2_{\mathrm{CBTZ}} .
\label{eq:metric-scaling}
\ee
In a time-independent scaling of a static metric, the temperature
scales as inverse distance, as seen from the period of thermal Green's
functions. If this scaling is assumed to hold also in our
time-dependent situation, equations \nr{T-BTZ} and
\nr{eq:metric-scaling} lead again to the time-dependent temperature and
entropy density~\nr{T-s}, the entropy density now being fixed by the
proportionality argument $s\sim T$.

\end{itemize}

Finally, recall that the gravity side of the duality is not
just AdS$_3$ but AdS$_3\times$S$_3\times T_4$. The parameter
$\CL/G_3=4Q_1Q_5$ is fixed by~\nr{g3}, and we arrive at the final
result
\be
\epsilon(T)=p(T)=
\pi Q_1Q_5 T^2,
\qquad
s(T)=2\pi Q_1Q_5T.
\label{ideal+}
\ee
We may compare this with $(1+1)$-dimensional ideal gas in thermal
equilibrium with $N_b=4Q_1Q_5$ bosonic and $N_f=4Q_1Q_5$ fermionic
massless degrees of freedom, for which
\be
\epsilon(T)=p(T)=(N_b+\fr12 N_f){\pi\over6}T^2=
\pi Q_1Q_5T^2,\qquad s(T)=2\pi Q_1Q_5T.
\label{idealgas}
\ee
The results duly coincide.
For AdS$_5\times$S$_5$ the boundary pressure can be computed in
the strong coupling limit $g^2N_c\gg1$. There its
value is 3/4 times the weakly coupled ideal gas value.

To end this section, we wish to compare our results for the boundary
thermodynamics to a method that was applied to the corresponding
$(3+1)$-dimensional situation in~\cite{jp}. Let us assume $M>1$ and
compare the time-dependent metric \nr{t-solution-M} and the static
metric~\nr{eq:static-metric}. The latter looks like the former if one
replaces $z_h=2\CL/\sqrt{M}$ by $z_h=v\tau$, where $v$ is given
by~\nr{v-def}, and also replaces $\CL^2d\eta^2$ by $\tau^2d\eta^2$.
This suggests that the static temperature $T_{\mathrm{BTZ}} = 1/(\pi
z_h)$ \nr{T-BTZ} should be replaced by the time-dependent temperature
\be
T
= \frac{1}{\pi z_h}
= \frac{1}{\pi v\tau}
= \frac{\sqrt{M-1}}{2\pi\tau} .
\label{T-from-moving}
\ee
For the entropy, we can derive a consistent result using the area formula
\be
S={A\over 4G_3},\qquad A = \int\,d\eta\sqrt{\gamma(z_h,\tau)}=
\int\,d\eta{\CL\over z_h}2\tau={2\CL\over v\tau}\int \tau d\eta,
\ee
where $\gamma$ is the determinant of the metric on the one-dimensional
hypersurface of constant $\tau$ and~$z$. To convert $S$ to an entropy
density, we have to divide by the volume, which again is the standard
longitudinal boost invariant expression $\int \tau d\eta$. The
entropy density thus becomes
\be
s={\CL\over 2G_3}\,{1\over v\tau}
= {\CL\over 4G_3}\,{\sqrt{M-1}\over \tau}.
\label{s-from-moving}
\ee
Using formula \nr{eps} for the energy density and the pressure, we find
\be
\epsilon(\tau)=p(\tau)
={\pi\CL\over 4 G_3}
\left( 1 - \frac{1}{M} \right)^{-1}
T^2(\tau),
\quad
s(\tau)={\pi\CL\over 2 G_3}T(\tau).
\ee
The proper thermodynamic relation $s(T)=p'(T)$ is thus recovered in
the semiclassical limit, $M\gg1$. Alternatively, if
we view the total boundary energy-momentum tensor \nr{raw-T} as
coming from a fluid, without a vacuum component, formula
\nr{eps} is modified
by the replacement $M \to M-1$, and the thermodynamic relation
$s(T)=p'(T)$ is recovered from \nr{T-from-moving} and
\nr{s-from-moving}
for all $M>1$.

The result from our method for the temperature and entropy density
is given by Eq.\nr{T-s} and holds for $M\ge0$. This
result and the method of
\cite{jp} are thus in agreement in the semiclassical limit,
$M\gg1$. In this limit the temperature satisfies
$\pi T\tau=\sqrt{M}/2 \gg 1$ and is
therefore consistent with the uncertainty principle.
We view this agreement as indirect support
for the semiclassical limit of the
thermodynamical conclusions obtained
in~\cite{jp}, where the global structure
of the $(4+1)$-dimensional bulk spacetime remained open.
We emphasise, however, that our method is coordinate-invariant:
the bulk solution
\nr{t-solution-M} defines a specific way of extending the
boost isometry on the boundary into a spacelike translational
isometry in the bulk, and in our method
the thermodynamics arises from the
Killing horizon of the timelike Killing vector that commutes with
these spacelike translations. From the geometric viewpoint, we
therefore view our method as more reliable beyond the
semiclassical limit, at least in the regime in which
a classical bulk
solution without quantum corrections can be expected to
give accurate results for the boundary quantum
theory~\cite{malda-stro,louko-marolf}.

\section{Bulk metric with rotation}
\label{sec:gen-BTZ}

If the starting ansatz \nr{ansatz} is generalised to include a term
proportional to $d\tau d\eta$, integration of the field equations
with our boundary condition \nr{milne-metric} yields one more constant
of integration, corresponding to the rotation of the black hole. We
shall now find this rotating generalisation of
\nr{t-solution-M} by working backwards from the
rotating BTZ metric \cite{henneaux,carlip},
\be
ds^2 =
- F dt^2
+ \frac{dr^2}{F}
+ r^2 {(d\varphi + N^\varphi dt)}^2 ,
\label{gen-BTZ}
\ee
where
\be
F =
\frac{(r^2 - r_+^2) (r^2 - r_-^2)}{\CL^2 r^2} ,
\qquad
N^\varphi =
- \frac{r_+ r_-}{\CL r^2} .
\ee
This metric satisfies Einstein's equations \nr{einstein-equations} and
is therefore locally isometric to AdS$_3$.
The parameters $r_\pm$ determine the mass parameter $M$ and the
angular momentum parameter $J$ by
\be
M =
\frac{r_+^2 + r_-^2}{\CL^2} ,
\qquad
J =
\frac{2 r_+ r_-}{\CL} .
\ee

\begin{figure}[!tbp]
\begin{center}
\includegraphics[width=0.6\textwidth]{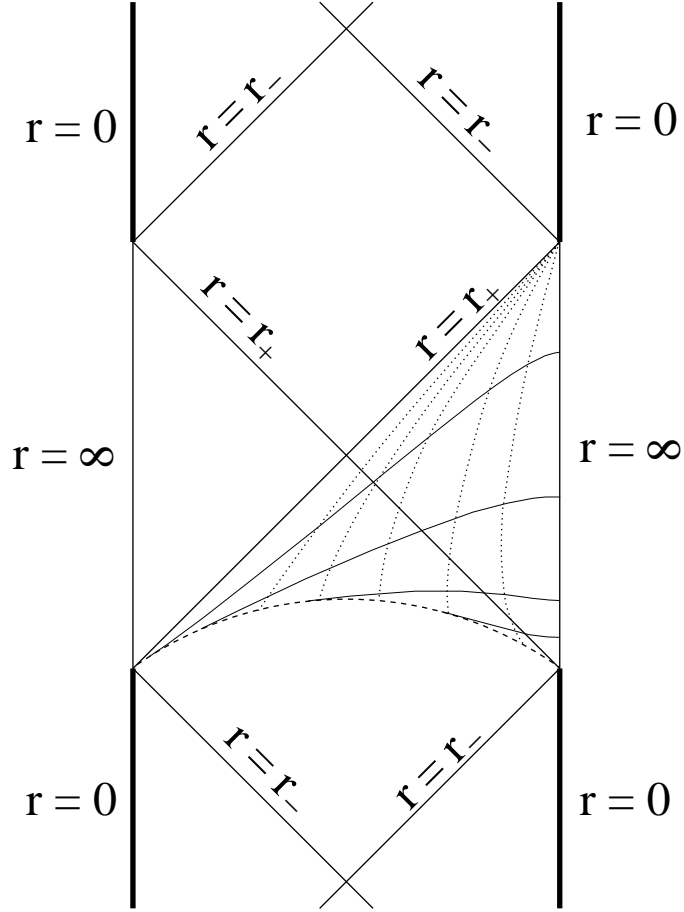}
\end{center}
\caption{\small
The conformal diagram of the
non-extremal BTZ black-and-white hole spacetime with $J\ne0$
\cite{henneaux,carlip}.
The solid (respectively dotted) lines are curves of constant $\tau$
(constant $z$) in the region $0 < z < v\tau$ of the
metric~\nr{tau-z-rot-sol}, where $M>1$, ${(M-1)}^2 - {(J/\CL)}^2 >0$
and $v$ is given by~\nr{vh-def}.
$\tau$~increases upwards and $z$ increases to the
left.
The dashed line is the coordinate singularity $z = v\tau$.}
\label{BTZspinfigure}
\end{figure}

When $\varphi$ is periodic with period $2\pi$ and the parameters
satisfy $0 \le |r_-| < r_+$, or equivalently $0 \le |J| <
\CL M$, the metric
\nr{gen-BTZ} is the nonextremal BTZ black hole. The
Boyer-Lindquist -type coordinates $(t,r,\varphi)$ are singular at
$r=r_+$, which is the Killing horizon of the Killing
vector~$\partial_t$, and if $r_-\ne0$, there is also a coordinate
singularity at $r=|r_-|$, which is an inner Killing horizon
of~$\partial_t$. Both of these horizons are nondenegenerate.  For
$r_-=0$ the metric reduces to~\nr{standard}, and the conformal diagram
was shown in Figure~\ref{BTZnospinfigure}. For $r_-\ne0$, the
conformal diagram is shown in Figure~\ref{BTZspinfigure}. The ADM
energy and the angular momentum at the infinity $r\to\infty$ are
respectively $M/(8G_3)$ and $J/(8G_3)$.
For us $\varphi$ is not periodic, but the crucial point is again that
the observations about the Killing vector $\partial_t$ remain valid
even for nonperiodic~$\varphi$, and the only Killing vectors that
commute with translations in $\varphi$ are $\partial_\varphi$
and~$\partial_t$. We shall continue to describe the metric in the
black hole terminology even for nonperiodic~$\varphi$.

For definiteness, assume for the moment that $M>1$ and ${(M-1)}^2 -
{(J/\CL)}^2 >0$, which is a subcase of the nondegenerate black hole
range $|J| <  \CL M$. Starting from the exterior $r>r_+$, we first
transform from $(t,r,\varphi)$ to $(\beta, r, \eta)$ by
\be
t = \CL [\beta - f(r)] ,
\qquad
\varphi =
\eta +
\CL \int^r N^\varphi(\tilde{r}) f'(\tilde{r}) d\tilde{r},
\label{rot-transf}
\ee
where $f$ is a function of $r$ only and satisfies
\be
f' = - \frac{1}{\CL F \sqrt{F+1}} .
\label{fprime-def}
\ee
The metric becomes
\be
ds^2 =
- \CL^2 (F+1) d\beta^2
+ \CL^2 \left( d\beta - \frac{dr}{\CL\sqrt{F+1}} \right)^2
+ r^2 {\left(d\eta + \CL N^\varphi
d\beta \right)}^2 .
\label{BTZ-F}
\ee
We then define $\alpha$ as the positive solution of
\be
{(r/\CL)}^2
=
\frac{e^{2\alpha} + e^{-2\alpha}}{v^2}
+ \frac{M-1}{2} ,
\label{rsquared-alpha}
\ee
where
\be
v = \frac{2}{\left[{(M-1)}^2 - {(J/\CL)}^2\right]^{1/4}} .
\label{vh-def}
\ee
It follows that
\be
\frac{dr}{\CL\sqrt{F+1}} = d\alpha ,
\label{dr/dalpha}
\ee
\be
F+1 =
\frac{\left( e^{2\alpha} - e^{-2\alpha} \right)^2}
{v^2 \left[ e^{2\alpha} + e^{-2\alpha}
+ \frac12 v^2 (M-1)\right]} .
\label{F+1}
\ee
Substituting
\nr{rsquared-alpha},
\nr{dr/dalpha}
and
\nr{F+1}
in
\nr{BTZ-F}
yields in the coordinates $(\beta,\alpha,\eta)$ a metric that is
regular for $\alpha>0$, rather than just for $r> r_+$. Finally, defining
the coordinates $(\tau,\eta,z)$ by
\be
z/\CL = v e^{\beta -\alpha},\qquad \tau/\CL = e^\beta
\ ,
\label{gen-zttau-alphabeta}
\ee
the metric takes the form
\ba
ds^2 &=& \frac{\CL^2}{z^2}
\left(
-
\left\{
1 - \frac{(M-1) z^2}{2 \tau^2}
+ \frac{1}{16}
\left[{(M-1)}^2 - {(J/\CL)}^2\right]
\frac{z^4}{\tau^4}
\right\} d\tau^2
- \frac{J z^2}{\CL \tau} d\tau \, d\eta
\right.
\nonumber
\\
\noalign{\medskip}
&&
\hspace{5ex}
\left.
+
\left\{
1 + \frac{(M-1) z^2}{2 \tau^2}
+ \frac{1}{16}
\left[{(M-1)}^2 - {(J/\CL)}^2\right]
\frac{z^4}{\tau^4}
\right\}
\tau^2 d\eta^2
+ dz^2
\right) .
\label{tau-z-rot-sol}
\ea
This is the promised rotating generalisation
of~\nr{t-solution-M}, reducing to \nr{t-solution-M} for $J=0$.

Although the transformations shown above assumed $M>1$ and ${(M-1)}^2
- {(J/\CL)}^2 >0$, an analytic continuation argument (or an explicit
computation) shows that
\nr{tau-z-rot-sol} solves Einstein's equations
\nr{einstein-equations} for any values of $M$ and~$J$.
One of the Killing vectors is~$\partial_\eta$, and the metric has at
$z\to0$ the asymptotically AdS form of \nr{AdS-form-in-z} and
\nr{gexp} with the conformal boundary metric~\nr{milne-metric}. The
transformation to the BTZ form
\nr{gen-BTZ} can be found for any $M$ and $J$ by a suitable
generalisation of the above formulas.

When $0 < |J| < \CL M$, so that the spacetime is a nondegenerate black
hole, the region covered by the metric \nr{tau-z-rot-sol} can be found
as in section~\ref{sec:BTZ-spinless}. Consider in particular the case
in which $M>1$ and ${(M-1)}^2 - {(J/\CL)}^2 >0$. The coordinate
singularity in
\nr{tau-z-rot-sol} is then at $z = v\tau$, and a regular region that is
asymptotically AdS at $z\to0$ is $0< z < v\tau$, corresponding to
$\alpha>0$ in the coordinates $(\beta,\alpha,\eta)$. Examination of
the coordinate transformations given above and the properties of the
extended BTZ spacetime \cite{henneaux,carlip} shows that the situation
is qualitatively similar to that with $J=0$. The coordinate singularity is
again on a spacelike surface in the white hole region, and this
coordinate singularity is neither an event horizon, apparent horizon
nor a dynamical horizon
\cite{hawking-ellis,ashtekar-dynamical}. The curves of constant $\tau$
and the curves of constant $z$ are shown in the conformal diagram in
Figure~\ref{BTZspinfigure}.

\section{Energy-momentum from the rotating bulk}
\label{sec:rot-energymomentum}

The boundary energy-momentum tensor for the bulk
metric~\nr{tau-z-rot-sol}, with arbitrary values of $M$ and~$J$, can
be computed directly from~\nr{skende}.  In the coordinates
$(\tau,\eta)$, the boundary metric is the Milne metric
\nr{milne-metric} and we obtain
\be
T_{\mu \nu} =
{\CL\over 16\pi G_3 \tau^2}
\left(
\begin{array}{cc}
M-1 & -(J/\CL)\tau \\
\noalign{\smallskip}
-(J/\CL)\tau & (M-1)\tau^2
\end{array}
\right) .
\label{milne-T}
\ee
As the components in \nr{milne-T} do not depend on~$\eta$, the
energy-momentum tensor is invariant under the boosts generated by the
boundary Killing vector~$\partial_\eta$. This had to be the case since
$\partial_\eta$ is a Killing vector also in the bulk.


As in section~\ref{sec:thermodynamics}, we split the energy-momentum
tensor as $T_{\mu \nu} = T_{\mu\nu}^{(\mathrm{fluid})} +
T_{\mu\nu}^{(\mathrm{vac})}$, where the vacuum contribution
$T_{\mu\nu}^{(\mathrm{vac})}$ \nr{vac-T} is what remains in the limit
of vanishing $M$ and $J$ and the fluid contribution
$T_{\mu\nu}^{(\mathrm{fluid})}$ is
\be
T_{\mu\nu}^{(\mathrm{fluid})} =
{1\over 16\pi G_3 \tau^2}
\left(
\begin{array}{cc}
\CL M & -J\tau \\
\noalign{\smallskip}
-J\tau & \CL M\tau^2
\end{array}
\right) .
\label{J-fluid-T}
\ee
A nonzero $J$ clearly affects $T_{\mu\nu}^{(\mathrm{fluid})}$. We wish
to understand how.

When $|J| < \CL |M|$, $T_{\mu\nu}^{(\mathrm{fluid})}$ \nr{J-fluid-T}
can be written in the perfect fluid form~\nr{milne-T-perffluid}, where
the normalised velocity vector of the fluid is
\ba
u^\mu
&=& (\cosh\phi , \tau^{-1} \sinh\phi) ,
\label{u-def}
\ea
$\phi$ is given by
\be
\tanh2\phi = \frac{J}{\CL M} ,
\ee
and the energy density $\epsilon$ and pressure $p$ of the fluid in its
local rest frame are
\be
\epsilon=p=
\frac{\CL M}{16\pi G_3 \tau^2}
\sqrt{1- \left(\frac{J}{\CL M}\right)^2} .
\label{eps-rot}
\ee
As the components in \nr{u-def} do not depend on~$\eta$, the
velocity vector field $u^\mu$ is invariant
under the boosts generated by~$\partial_\eta$, and $u^\mu$ has at each
point the peculiar velocity $v_\rmi{pec}=\tanh\phi$
relative to the flow field $\partial_\tau$ of the co-moving Milne
observers.
From the gauge theory viewpoint, the boundary energy-momentum tensor
therefore describes a perfect fluid flow that is invariant under the
longitudinal boosts and is at each point moving with respect to the
Bjorken similarity flow
\nr{simflow} with the longitudinal velocity~$v_\rmi{pec}$.
Note, however, that the flow lines are
not inertial when $\phi\ne0$. In the Minkowski null coordinates
$(x^+,x^-)$, in which
$x^{\pm} = \tau e^{\pm\eta}/\sqrt{2}$ and
$ds^2 = -2 dx^+ \, dx^-$, we have
\be
u^+ = e^\phi \sqrt{\frac{x^+}{2x^-}} , \qquad
u^- = e^{-\phi} \sqrt{\frac{x^-}{2x^+}} ,
\ee
and the flow lines are
\be
x^+ = K {(x^-)}^{\exp(2\phi)} ,
\label{eq:rot-flow}
\ee
where the positive constant $K$ labels the lines.
It can be verified that on each line the
proper time $\lambda$ is given by $\lambda = \sqrt{2 x^+ x^-}/
\cosh(\phi)$ and the proper acceleration has the magnitude
$|\tanh\phi|/\lambda$.
While all the
flow lines start from
$(x^+,x^-)= (0,0)$ at $\lambda=0$, the proper
acceleration diverges as $\lambda\to0$, and although the proper
acceleration approaches zero as $\lambda\to\infty$, the fall-off is so
slow that the flow lines are not asymptotically inertial at
$\lambda\to\infty$.


If, in addition to $|J| < \CL |M|$, we assume also $M>0$,
the bulk solution is a black hole, and we can use its thermodynamics
to equip the flow
\nr{eq:rot-flow} with a temperature and an entropy density as in
section~\ref{sec:thermodynamics}. It remains however an open problem
to identify a microscopic gauge theory process whose macroscopic
properties the flow
\nr{eq:rot-flow} and its associated thermodynamical quantities
would describe.

For completeness, consider briefly the remaining ranges of the
parameters.
When
$|J| = \CL |M| \ne0$, the bulk black hole is extremal and
$T_{\mu\nu}^{(\mathrm{fluid})}$
takes the null dust form
\be
T_{\mu\nu}^{(\mathrm{fluid})} =
\frac{\CL M}{16\pi G_3 \tau^2}
k_\mu k_\nu ,
\label{milne-T-nulldust}
\ee
where $k^\mu = \left(1 , \pm \tau^{-1}  \right)$ is a null vector
and the sign in the second component is that of $\CL M /J$.  When $J =
\CL M =0$, $T_{\mu\nu}^{(\mathrm{fluid})}$ vanishes. When $\CL M <
|J|$, $T_{\mu\nu}^{(\mathrm{fluid})}$ has no real eigenvectors and
does not arise from conventional matter
fields~\cite{hawking-ellis}. We conclude that
$T_{\mu\nu}^{(\mathrm{fluid})}$ is that of a perfect fluid with a
positive energy density precisely when $|J| < \CL |M|$, that is,
precisely when the bulk solution is a nondegenerate black hole.

\section{Quasinormal modes}

As the metric \nr{t-solution-M} with $M>0$ does not cover the black
hole horizon, but does cover part of the white hole region, it is
tempting to interpret the ion collision on the boundary as dual to an
eruption from a white hole in the bulk. We emphasise that our
derivation of the boundary temperature and entropy density \nr{T-s}
did
\emph{not\/} rely on such an interpretation
but operated directly on the Killing horizon of the extended bulk
solution and its thermodynamical properties. From a general
relativistic viewpoint, one is indeed inclined to read little into the
behaviour of a specific set of coordinates in the deep bulk region,
even when the coordinates are near the boundary adapted to the
boundary physics of interest: it is always possible to introduce a
smooth coordinate transformation that is the identity in a
neighbourhood of the boundary but drastically changes the region
covered deep in the bulk. We show in the Appendix that coordinates
similar to those in
\nr{t-solution-M} can be introduced even in $(1+1)$-dimensional
Minkowski spacetime, with the coordinate singularity on a spacelike
curve in the past light cone of the origin.

All that being said, suppose one did wish to adopt the eruption from a
white hole as a serious physical picture in the bulk. Would this
picture have any observable consequences in the boundary physics? We
shall now argue that the bulk quasinormal modes
\cite{birmingham-quasi,bir-sachs-solo,son-starinets}
may provide insight into this question.

For concreteness, we consider a bulk scalar field $\Phi$ that
satisfies the Klein-Gordon equation
\be
\left( \Box - \frac{\mu}{\CL^2} \right) \Phi =0 ,
\label{KG-eq}
\ee
where $\mu>0$. For technical
simplicity (cf.~\cite{son-starinets}), we further assume that
$\sqrt{1+\mu}$ is not an integer.

Consider the quadrant $U<0$, $V>0$ of the spinless BTZ spacetime
\nr{spinlessBTZ-kruskal} in the Schwarzschild-like
coordinates~\nr{standard}. We look for a solution to \nr{KG-eq} that
is independent of $\eta$ (corresponding to boost invariance on the
boundary) and has the separable form $\Phi = e^{-i\omega t} R(r)$,
where $\omega$ is a nonvanishing complex number. A~pair of linearly
independent solutions for $R(r)$ is \cite{birmingham-quasi}
\be
R_\pm(r) = \left( 1 - \frac{r_+^2}{r^2} \right)^{\nu_\pm}
\left( \frac{r_+^2}{r^2}\right)^{\gamma}
F
\left(
\nu_\pm + \gamma, \, \nu_\pm + \gamma ; \,
2\nu_\pm + 1; \, 1 - \frac{r_+^2}{r^2}
\right) ,
\ee
where
\ba
\nu_\pm &=& \pm \frac{i \CL^2 \omega}{2 r_+} ,
\label{nu-def}
\\
\gamma &=& \frac12 \left( 1 - \sqrt{1+\mu} \right) ,
\ea
$F$ is the hypergeometric function,
and we are assuming that $2\nu_\pm+1$ is not a negative integer.
We write $\Phi_\pm(t,r) = e^{-i\omega t} R_\pm(r)$.
Using
\nr{kruskal-to-schw} to go to the Kruskal
coordinates~\nr{spinlessBTZ-kruskal}, we see that $\Phi_+$ continues
regularly across the past branch of the horizon at $r=r_+$, into the
white hole region of the spacetime, but it is singular on the future
branch of the horizon and cannot thus be regularly continued into the
black hole region. Conversely, $\Phi_-$ continues regularly into the
black hole region but not into the white hole region.

We now require $\Phi_\pm$ to vanish at $r\to\infty$. Using equation
15.3.6 in~\cite{abramowitz}, and the technical assumption that
$\sqrt{1+\mu}$ is not an integer, we find that this happens precisely
when $\nu_\pm = -(n+1) + \gamma$, where $n = 0,1,\ldots$. Dropping an
overall constant, the solutions then take the form
\ba
\Phi_{\pm,n}(t,r) &=&
\exp \! \left[ \pm \frac{2 r_+ (n+1-\gamma) t}{\CL^2}
\right]
\left( \frac{r_+^2}{r^2}\right)^{1 -\gamma}
\left( 1 - \frac{r_+^2}{r^2} \right)^{\gamma - (n+1)}
\nonumber
\\
&&
\times \,
F
\left(
-n, \, -n ; \, 2(1-\gamma); \, \frac{r_+^2}{r^2}
\right) ,
\qquad
n =  0,1,\ldots \, .
\ea
$\Phi_{-,n}$ are the usual bulk quasinormal
modes~\cite{birmingham-quasi}, decaying in $t$ by falling into the
black hole. These modes are singular at the white hole horizon, but
this singularity in the past does not pose a problem for using
$\Phi_{-,n}$ to describe decay processes whose initial conditions are
set in the exterior. By contrast, the time-reversed bulk modes
$\Phi_{+,n}$ erupt from the white hole and become singular on reaching
the black hole horizon. Because of this singularity, $\Phi_{+,n}$ are
not usually considered relevant for bulk physics whose initial
conditions are set in the exterior. It seems however less clear
whether this singularity would preclude $\Phi_{+,n}$ from describing
physics that takes place on the boundary, since the black hole horizon
is not in the causal past of the boundary.

Thus, if the eruption from a white hole is proposed to have a physical
meaning as the dual to the boundary ion collision, a possible
consequence is that the relevant solutions to the wave equation
\nr{KG-eq} should be the eruption modes, rather than the usual
quasinormal modes. From
\nr{tau-z-to-kruskal} and \nr{kruskal-to-schw} we find that
$\Phi_{\pm,n}$ have at $z\to0$
the asymptotic form
\be
\Phi_{\pm,n} \sim
\left(\frac{\tau}{z}\right)^{2(\gamma-1)}
\left(\frac{\tau}{\CL}\right)^{\pm2(r_+/\CL)(n+1-\gamma)} .
\label{Phi-as}
\ee
The quasinormal modes therefore have on the boundary a decreasing
power-law behaviour in~$\tau$, as one expects of a relaxation
process. However, the eruption modes have an
\emph{increasing\/} power-law behaviour in $\tau$ for large~$n$,
and even for all $n$ if the hole is so large that $r_+/\CL >1$.

Finding a power-law increase in some thermodynamical variables on the
boundary would thus provide smoking-gun evidence for the eruption
picture in the bulk, and it would also suggest a similar picture in
the physically more interesting system with a $(3+1)$-dimensional
boundary~\cite{jp}. Conversely, the absence of a power-law increase in
the boundary termodynamics would discourage the eruption picture as
one with physical consequences for the boundary. We leave further
scrutiny of this question to future work.

\section{Conclusions}

Gauge theory/gravity duality has made interesting predictions about
matter in static thermal equilibrium \cite{gkp,pss}. An obvious
problem is to explore what, if anything, the same framework can say
about systems in expansion.

In this paper we have studied this problem for
$(1+1)$-dimensional matter that expands as the Bjorken similarity
flow. Using an ansatz \nr{ansatz} adapted to the symmetries
we found an explicit time dependent AdS$_3$ bulk solution \nr{t-solution-M}
with a time dependent coordinate singularity in
the bulk. By explicit coordinate transformations we showed that
the solution is part of the spinless BTZ
black hole, with the angular dimension unwrapped. The coordinate
singularity is on a spacelike surface in the white hole region.

The holographic energy-momentum tensor on the boundary was
computed with standard techniques and separated into two components,
a vacuum contribution (coming from the bulk metric with $M=0$)
and a contribution that corresponds to the boundary fluid.
This separation turned out important for a consistent definition of the
temperature and entropy density of the fluid.
The fluid was found to
be an ideal gas in adiabatic expansion. We also recovered
the known fact that the
boundary pressure, as calculated in the AdS$_3\times$S$_3\times T_4$
approach, is exactly the same as that of a massless ideal gas of an
appropriate number of bosons and fermions in one spatial dimension.

To obtain the time-dependent temperature and
entropy density of the fluid, we first computed the
time-independent temperature and entropy density
of the BTZ hole, defined in a standard way
with respect to the longitudinal volume element
$\CL d\eta$, and we then performed the time-dependent scaling
$\CL\leftrightarrow\tau$ to the longitudinal volume element
$\tau d\eta$ that is appropriate for the boost symmetry of
the Bjorken flow on the boundary. We emphasise that
the thermodynamic results therefore relied in no way on the
coordinate singularity in the metric~\nr{t-solution-M}.
In a more geometric language, the bulk solution
\nr{t-solution-M} defines a specific way of extending the
boost isometry on the boundary into a spacelike translational
isometry in the bulk, and the thermodynamics arose from the
Killing horizon of the timelike Killing vector that commutes with
these spacelike translations.

Our AdS$_3$ solution, of course, does not contain
the full dynamical
content of the physically interesting
case where the boundary is $(3+1)$-dimensional.
For example, in our case $\epsilon(\tau)$ is proportional to
$\tau^{-2}$ for all~$\tau$, while on the
$(3+1)$-dimensional boundary $\epsilon(\tau)$ appears to
encode qualitatively different physics at large \cite{jp}
and small \cite{kovchegov}~$\tau$.
However, in our case the global structure
of the bulk metric is completely known.
We were in particular able to verify that our
coordinate-invariant derivation of the thermodynamics
was in the semiclassical regime ($M\gg1$) fully compatible with the
method of~\cite{jp}, in which
the temperature and entropy formulas of a static black hole
are extended to the time-dependent case by comparing the
coordinate singularities.

We also showed that inclusion of angular momentum in the bulk leads to a
boundary flow that is still boost invariant but has a nonzero
longitudinal velocity with respect to the Bjorken expansion. Finally,
we argued that the bulk quasinormal modes may shed light on the
possible physical relevance, or lack thereof, of the observation that
the coordinates in \nr{t-solution-M} cover part of the bulk white hole
interior but none of the black hole interior.

The boundary flow temperature \nr{T-s} can be written in the form
\be
T(\tau)=T_i{\tau_i\over\tau},\quad \pi T_i\tau_i=\fr12\sqrt{M}.
\label{tfinal-2}
\ee
Interesting physical questions here are what is the smallest time
$\tau_i$ for which \nr{tfinal-2} holds, what is the thermalisation
time, and what is the associated maximum temperature. To address these
questions, additional information about the process would need to be
introduced. This information could regulate the singularity in
\nr{tfinal-2} by $1/\tau\to1/(\tau+\tau_0)$ or produce other
corrections, for example of the type $1/\tau\to
1/\tau-\tau_0/\tau^2$. As the gauge theory/gravity duality approach is
designed to give vacuum expectation values of gauge theory operators,
it is by no means obvious how it should be used for processes starting
at some $\tau=0$.

\vspace{0.5cm}
Acknowledgements. We thank Esko Keski-Vakkuri, D.~Podolsky, Eric Poisson
and Jung-Tay Yee for
discussions. This research has been supported by Academy of
Finland, contract number 109720 and by PPARC grant PP/D507358/1.
JL~thanks Helsinki Institute of Physics
for hospitality during the final stages of the work.

\begin{appendix}

\section*{Appendix: Singular coordinates in $(1+1)$ Minkowski}

In this appendix we give an example of a coordinate system in
$(1+1)$-dimensional Minkowski spacetime with a singularity structure
similar to that in~\nr{t-solution-M}.

Let $(T,X)$ be the usual Minkowski coordinates, $ds^2 = - dT^2 +
dX^2$. We introduce the coordinates $(\tau,z)$ by
\ba
T-X &=&
-
\frac{\ln(v\tau/z) \, +1}{\tau} ,
\nonumber
\\
\noalign{\smallskip}
T+X &=&
\left[ \ln(v\tau/z) \, -1 \right] \tau ,
\ea
where $0<z<\infty$, $0<\tau<\infty$, and $v$ is a positive
constant. The metric becomes
\be
ds^2 =
\frac{1}{z^2}
\left\{
- {\left[ \left(\frac{z}{\tau}\right)
\ln \! \left(\frac{v\tau}{z}\right)\right]}^2 d\tau^2
+ dz^2
\right\} .
\label{eq:Mink-peculiar}
\ee
This metric is regular for $0<z<v\tau$ and $0<v\tau<z$, and the
oordinate singularity at $z=v\tau$ is on the spacelike curve $T = -
\sqrt{X^2 + 1}$. The curves of constant $\tau$ and
the curves of constant $z$ in the region $0<z<v\tau$ are qualitatively
similar to those shown in Figure~\ref{BTZnospinfigure}.

\end{appendix}

\end{document}